\documentclass[twocolumn,showpacs,superscriptaddress,preprintnumbers,prd]{revtex4}
\usepackage{graphicx,amsmath,dcolumn,bm}
\usepackage{feynmp}
\usepackage{epic,eepic}
\usepackage[dvipdfm]{pict2e}
\newcommand{\ra}{\rightarrow}
\def\lsim{\mathrel{\rlap{\lower4pt\hbox{\hskip1pt$\sim$}}
    \raise1pt\hbox{$<$}}}
\def\slash#1{\hspace*{-0.05cm}\not\!#1}

\usepackage{slashed}
\newcommand{\pslash}{\slashed{p}}
\begin{document}
%%%%%%%%%%%%%%%%%%%%%%%%%%%%%%%%%%%%%%%%%%%%%%%%%%%%%%%%%%%%%%%%%%%%%%%%%%%%%%%%
\preprint{KEK-TH-1623, J-PARC-TH-0026}
\title{Determination of exotic hadron structure \\
by constituent-counting rule for hard exclusive processes}
\author{H. Kawamura}
\affiliation{KEK Theory Center,
             Institute of Particle and Nuclear Studies \\
             High Energy Accelerator Research Organization (KEK) \\
             1-1, Ooho, Tsukuba, Ibaraki, 305-0801, Japan}
\author{S. Kumano}
\affiliation{KEK Theory Center,
             Institute of Particle and Nuclear Studies \\
             High Energy Accelerator Research Organization (KEK) \\
             1-1, Ooho, Tsukuba, Ibaraki, 305-0801, Japan}
\affiliation{
             J-PARC Branch, KEK Theory Center,
             Institute of Particle and Nuclear Studies, KEK \\
%             High Energy Accelerator Research Organization (KEK) \\
           and
           Theory Group, Particle and Nuclear Physics Division, 
           J-PARC Center \\
           203-1, Shirakata, Tokai, Ibaraki, 319-1106, Japan}
\author{T. Sekihara}
\affiliation{KEK Theory Center,
             Institute of Particle and Nuclear Studies \\
             High Energy Accelerator Research Organization (KEK) \\
             1-1, Ooho, Tsukuba, Ibaraki, 305-0801, Japan}
           \date{\today}
%\date{July 22, 2013}
%\date{June 29, 2013}
%%%%%%%%%%%%%%%%%%%%%%%%%%%%%%%%%%%%%%%%%%%%%%%%%%%%%%%%%%%%%%%%%%%%%%%%%%%%%%%%
\begin{abstract}
We propose to use hard exclusive production of an exotic hadron
for finding its internal quark-gluon configuration 
by the constituent-counting rule 
in perturbative QCD. In particular, the cross section 
for the exclusive process $\pi^- + p \to K^0 + \Lambda (1405)$ 
is estimated at the scattering angle $\theta=90^\circ$
in the center-of-mass frame by using current experimental data.
In comparison, the cross section for the ground-state
$\Lambda$ production $\pi^- + p \to K^0 + \Lambda$ is also shown.
We suggest that the internal quark configuration of $\Lambda (1405)$
should be determined by the asymptotic scaling behavior of 
the cross section. If it is an ordinary three-quark baryon,
the scaling of the cross section is $s^{8} d\sigma /dt=$constant, whereas
it is $s^{10} d\sigma /dt=$constant if $\Lambda (1405)$
is a five-quark hadron, where $s$ and $t$ are Mandelstam variables.
Such a measurement will be possible, for example, by using
the high-momentum beamline at J-PARC.
In addition, another exclusive process $\gamma + p \to K^+ + \Lambda (1405)$
could be investigated at LEPS and JLab for finding the nature
of $\Lambda \, (1405)$.
We indicate that the constituent-counting rule could be used as
a valuable quantity in determining internal structure of exotic hadrons
by high-energy exclusive processes, where quark-gluon degrees of freedom 
explicitly appear.
Furthermore, it is interesting to investigate the transition
from hadron degrees of freedom to quark-gluon ones for exclusive
exotic-hadron production processes.
\end{abstract}
%%%%%%%%%%%%%%%%%%%%%%%%%%%%%%%%%%%%%%%%%%%%%%%%%%%%%%%%%%%%%%%%%%%%%%%%%%%%%%%%
\maketitle

%%%%%%%%%%%%%%%%%%%%%%%%%%%%%%%%%%%%%%%%%%%%%%%%%%%%%%%%%%%%%%%%%%%%%%%%%%%%%%%%
%%%%%%%%%%%%%%%%%%%%%%%%%%%%%%%%%%%%%%%%%%%%%%%%%%%%%%%%%%%%%%%%%%%%%%%%%%%%%%%%
\section{Introduction}\label{intro}

A basic quark model indicates that baryons consist of three quarks
($qqq$) and mesons of a quark-antiquark pair ($q\bar q$). 
The family of the baryons and mesons is called hadrons,
and a few hundred hadrons have been found experimentally \cite{pdg-2012}.
However, an undoubted evidence has not been found for an exotic hadron, 
which has a different configuration from $qqq$ and $q\bar q$, 
although the fundamental theory of strong interaction, 
quantum chromodynamics (QCD), does not prohibit
the existence of such states like tetra-quark ($qq\bar q\bar q$)
and penta-quark ($qqqq\bar q$) hadrons \cite{jaffe-2005}.

It is, nevertheless, fortunate that the experimental situation changed
in the last several years because there have been reports 
on exotic hadron candidates particularly
from the Belle and BaBar collaborations \cite{belle-babar}. 
Exotic hadrons were suggested in experimental measurements so far
by looking at masses, spins, and decay widths, namely global observables 
at low energies. For example, electromagnetic and strong decay widths could
provide useful information on exotic hadrons \cite{exotic-decays}.
However, at low energies, effective degrees of freedom are
hadrons and only integrated quantities are observed,
so that it is not easy to judge whether or not a hadron has 
an exotic quark-gluon configuration. Therefore, it is appropriate
to look for high-energy processes, where quark-gluon degrees of 
freedom appear explicitly. Keeping this idea
in mind, we have been investigating possible high-energy processes
for determining internal structure of exotic hadron candidates,
for example, by fragmentation functions \cite{hkos08} and 
hadron-production processes in the $e^+ e^-$ annihilation \cite{hk2013}.

In this article, we propose that the constituent-counting rule 
of perturbative QCD could be used for finding the internal quark
configuration of exotic-hadron candidates in exclusive 
production processes. In the exclusive process $a+b \to c+d$
with large-momentum transfer, hard gluon exchange processes 
should occur to maintain the exclusive nature. Namely,
quarks should share large momenta so that they should
stick together to become a hadron by exchanging hard gluons.
Then, considering hard quark and gluon propagators 
in the reaction, we obtain that the cross section
of the $a+b \to c+d$ exclusive reaction should scale like
$d\sigma /dt \sim s^{2-n} f(\theta_{cm})$
with $n=n_a+n_b+n_c+n_d$,
where $s$ and $t$ are Mandelstam variables,
$\theta_{cm}$ is the scattering angle in the center-of-mass (c.m.) system,
and $n_h$ is the number of constituents in the particle $h$.
This asymptotic scaling relation is known as the constituent-counting rule 
\cite{counting-Matveev,counting-Brodsky,exclusive-Brodsky,counting-2,
Sivers-1982,Mueller-Brodsky}.
Since the factor $n_h$ clearly indicates the internal configuration
in the hadron, this scaling relation can be used for finding
internal configurations of exotic hadrons.

Here, we take an exotic hadron candidate $\Lambda (1405)$ as an
example for proposing such an idea.  The $\Lambda (1405)$ has been
controversial for many years from 1960's. The $\Lambda (1405)$ is a
baryon resonance with isospin 0, spin-parity $(1/2)^-$, strangeness
$-1$, mass 1405.1 MeV, and width 50 MeV \cite{pdg-2012}.
One of remarkable properties for $\Lambda (1405)$ is its anomalously
light mass.  Namely, in the ground states which possess spin-parity
$(1/2)^{+}$, $\Lambda$ is heavier than the nucleon, $M_{\Lambda} -
M_{N} \simeq + 180$ MeV, due to the heavier strange quark in $\Lambda$.  
However, in the $(1/2)^{-}$ states, the lowest excitation
states of $\Lambda$ and nucleon are $\Lambda (1405)$ and $N(1535)$,
respectively, and the puzzling reversal of the mass relation takes place
as $M_{\Lambda (1405)} - M_{N(1535)} \simeq - 130$ MeV, although
$\Lambda (1405)$ should have the heavier strange quark. 
The mass of $\Lambda (1405)$ is found to be anomalously light
also compared to the result of the SU(6) quark model,
in which both $\Lambda (1405)$ 
and $N(1535)$ should be considered to be baryons in the 70-dimensional
representation with $p$-wave excitation of a quark \cite{quark-model}
but it is difficult to explain the lighter mass of $\Lambda (1405)$ than
$N(1535)$ in the same representation.  Therefore, it has been thought
as an exotic hadron, beyond the naive three-quark ($uds$)
configuration.

Instead of an $uds$ three-quark system, the $\Lambda (1405)$ has been
considered as a $\bar K N$ molecule for a long time
\cite{dalitz} because it is slightly below the $\bar K N$ threshold
and the $\bar{K} N$ interaction is strongly attractive in the isospin
0 channel. There are recent theoretical progresses on $\Lambda
(1405)$ as a dynamically generated resonance in meson-baryon
scattering by the so-called chiral unitary model \cite{chiral-theory}. 
This model supports the meson-baryon molecule nature for
$\Lambda (1405)$ by revealing, {\it e.g.} predominance of the
meson-baryon component~\cite{Hyodo:2008xr}, its large-$N_{c}$ scaling
behavior~\cite{Hyodo:2007np}, and its spatial
size~\cite{sekihara-lambda-1405, Sekihara:2012xp}.  There is also a
proposal that $\Lambda (1405)$ could be a strange hybrid baryon by the
QCD sum rule \cite{hybrid}.  For the last several years, there are
many articles on $\Lambda (1405)$, so that we suggest the reader to
look at the reference section of the recent review
article~\cite{Hyodo:2011ur}. In the experimental side, precise
measurement of the $\Lambda (1405)$ line shape has been recently
performed in the photon induced $\Lambda (1405)$
production~\cite{lambda-1405-exp}, which provides information on
underlying dynamics and internal structure of $\Lambda
(1405)$~\cite{Nacher:1998mi}.  
In addition, hadron induced production experiments are currently in progress, 
{\it e.g.} by $pp$ collision at 3.5 GeV by
the HADES collaboration at 
GSI (Gesellschaft f\"ur Schwerionenforschung) \cite{Agakishiev:2012qx} 
and the $K^{-} + d$ reaction planned by the E31 experiment 
at J-PARC (Japan Proton Accelerator Research Complex)~\cite{Noumi:exp}.

In spite of theoretical studies on exotic hadrons for a long time, 
it is difficult to find a clear experimental evidence 
for the molecular or any other exotic configuration
because global quantities such as masses and decay widths
have been used. On the other hand, the quark-gluon degrees of freedom
appear in high-energy reactions. For example, scaling behavior of
exclusive cross sections is known as the constituent-counting rule.
In addition,
the transition from the hadron degrees of freedom to the quark-gluon ones 
seems to be clearly shown in the JLab measurements of
$\gamma + p \to \pi^+ + n$ \cite{jlab-gamma-exclusive}
by the differential cross section as the function 
of the c.m. energy $\sqrt{s}$.

In the same way, hard exclusive production processes of $\Lambda (1405)$
could be valuable for finding its internal quark configuration
by looking at the scaling behavior of the cross section at high energies.
Furthermore, it is interesting to investigate the transition phenomena 
from hadron degrees of freedom to the quark-gluon ones for exotic hadrons.
Fortunately, the high-momentum beamline of the J-PARC will be built
in a few years and unseparated hadron (essentially pion) beam
with momentum up to 15-20 GeV will be available.
Then, the exclusive reaction $\pi^- + p \to K^0 + \Lambda (1405)$
will become experimentally possible in principle.
However, no theoretical study exists for estimating 
the cross section of $\pi^- + p \to K^0 + \Lambda (1405)$
at large-momentum transfer. As far as we are aware, even an idea 
does not exist for studying the internal quark configuration
of exotic hadron candidates by the constituent-counting rule \cite{hk2013}. 
This article should be the first attempt to investigate
such an idea by taking $\Lambda (1405)$ as an example.
Since there is no prior study, we do not intend to show precise
theoretical cross sections, which are not possible at this stage
in any case. Instead, we try to provide an order of magnitude estimate 
of the cross sections in this work
for future experimental proposal at J-PARC or 
any other hadron facilities.

This article is organized in the following way.
In Sec. \ref{counting}, the constituent-counting rule is explained
for understanding cross-section behavior at high energies.
The cross sections are estimated in a high-momentum transfer region,
where the counting rule could be applied, by using existing measurements
for $\pi^- + p \to K^0 + \Lambda$ and
and $\pi^- + p \to K^0 + \Lambda (1405)$
in Sec. \ref{results}. 
We summarize our studies in Sec. \ref{summary}.

%%%%%%%%%%%%%%%%%%%%%%%%%%%%%%%%%%%%%%%%%%%%%%%%%%%%%%%%%%%%%%%%%%%%%%%%%%%%%%%%
%%%%%%%%%%%%%%%%%%%%%%%%%%%%%%%%%%%%%%%%%%%%%%%%%%%%%%%%%%%%%%%%%%%%%%%%%%%%%%%%
\section{Constituent-counting rule in hard exclusive reactions}\label{counting}

\begin{figure}[t!]
\vspace{0.0cm}
\begin{center}
   \includegraphics[width=5.0cm]{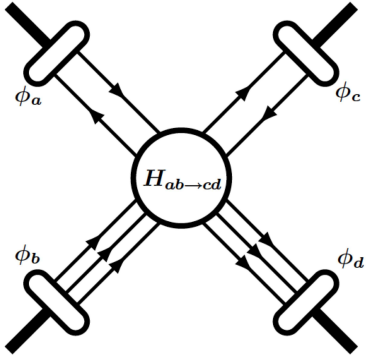}
\end{center}
\vspace{-0.4cm}
\caption{Hard exclusive scattering $a+b \to c+d$.}
\label{fig:elastic-ab-cd}
\end{figure}

We introduce the constituent-counting rule especially for the readers 
who are not familiar with perturbative QCD.
For a large-angle exclusive scattering $a+b \to c+d$,
the reaction cross section is given by
\begin{align}
\frac{d\sigma_{ab \to cd}}{dt}
% & = \frac{1}{16 \pi [ (s-m_a^2-m_b^2)^2 - 4m_a^2 m_b^2]}
& \simeq \frac{1}{16 \pi s^2}
\overline{\sum_{pol}} \, | M_{ab \to cd} |^2 ,
\label{eqn:two-body-cross}
\end{align}
where $s$ and $t$ are Mandelstam variables defined by
\begin{align}
& 
s = (p_a + p_b)^2 
% \simeq 2 p_a \cdot p_b 
    \simeq 4 \, | \, \vec p_{cm} \, |^2, 
\nonumber \\
& 
t = (p_a - p_c)^2 
% \simeq - 2 p_a \cdot p_c 
    \simeq -2 \, | \, \vec p_{cm} \, |^2 (1-\cos \theta_{cm}), 
\label{eqn:st}
\end{align}
where the masses of hadrons are neglected
by considering the kinematical condition $s, \, |t|\gg m_i^2$ ($i=a,b,c,d$),
$p_i$ is the momentum of the hadron $i$,
and $p_{cm}$ and $\theta_{cm}$ are momentum and scattering
angle in the c.m. frame, respectively.
Since we are considering the large-angle scattering, 
the kinematical invariants $s$ and $|t|$ are in the same order of 
magnitude. 
The summation of Eq. (\ref{eqn:two-body-cross})
indicates the average over the initial spins
and the summation for the final spins.
%%%%%
The matrix element $M_{ab \to cd} $ is expressed in the factorized form
at large momentum transfer
\cite{exclusive-Brodsky,Mueller-Brodsky,erbl}:
\begin{align}
&  
M_{ab \to cd} = \int [dx_a] \, [dx_b] \, [dx_c] \, [dx_d]  \,
    \phi_c ([x_c]) \, \phi_d ([x_d]) 
\nonumber \\
& 
\times 
H_{ab \to cd} ([x_a],[x_b],[x_c],[x_d],Q^2) \, 
      \phi_a ([x_a]) \, \phi_b ([x_b]) ,
\label{eqn:mab-cd}
\end{align}
in terms of the partonic scattering amplitude $H_{ab \to cd}$
and the light-cone distribution amplitude of each hadron,  
$\phi_a$, $\phi_b$, $\phi_c$, and $\phi_d$,   
as illustrated in Fig. \ref{fig:elastic-ab-cd}.
Here, $[x]$ indicates a set of the light-cone momentum 
fractions of partons in a hadron: $x_i=p_i^+/p^+$ where $p_i$ and
$p$ are $i$-th parton and hadron momenta, respectively,
and the light-cone component is defined as $p^+ = (p^0 + p^3)/\sqrt{2}$
by taking the third axis for the longitudinal direction.

%%%%%%%%%%%%%%%%%%%%%%%%%%%% figure %%%%%%%%%%%%%%%%%%%%%%%%%%%%
%%%%%%%%%%%%%%%%%%%%%%%%%%%% figure %%%%%%%%%%%%%%%%%%%%%%%%%%%%

For the nucleon, two independent variables are needed to describe
the distribution amplitude
by considering a constraint of the momentum conservation $x_1+x_2+x_3=1$.
Namely, we have $[x]=x_1, x_2$ and $[dx]=dx_1 dx_2$ in 
Eq. (\ref{eqn:mab-cd}). On the other hand, only one variable $x$
is needed for mesons such as pions and kaons (see Eq. (\ref{eqn:qbar-q-matrix}))
%%%%%%%
As an example, the reaction is illustrated in Fig. \ref{fig:elastic-ab-cd}
by taking the hadrons $a$ and $c$ as mesons and $b$ and $d$ as baryons.
The variable $Q^2$ indicates a hard scale of the reaction, which is 
given by $Q^2 \simeq s$ for the large-angle elastic exclusive 
scattering. 
In Eq. (\ref{eqn:mab-cd}), we have suppressed the renormalization and 
factorization scale dependencies: the former is controlled by 
the renormalization group equation for the coupling constant  
and the latter by the ERBL (Efremov-Radyushkin-Brodsky-Lepage)
evolution equations for the distribution 
amplitudes \cite{erbl}.  
Those scales are taken to be order of $Q^2$ to avoid large 
radiative corrections. The resulting $Q^2$ dependencies are logarithmic 
and do not largely affect the scaling behavior of the matrix element.     

%%%%%%%%%%%%%%%%%%%%%%%%%%%%%%%%%%%%%%%%%%%%%%%%%%%%%%%%%%%%%%%%%%%%%%%%%%%%%%%%
\subsection{Constituent-counting rule by dimensional counting}
\label{mass-counting}

Originally, the constituent-counting rule was suggested by dimensional 
counting \cite{counting-Matveev,counting-Brodsky}.
Then, it was studied by considering hard scattering processes 
in perturbative QCD \cite{counting-Brodsky,counting-2,exclusive-Brodsky}.
In this section, we explain derivation of the scaling rule
by counting mass dimensions. Then, we outline how the counting rule 
is understood in perturbative QCD in Sec. \ref{pqcd-counting}.

Because the state vector of a hadron is normalized as
$\langle \, h(p') \, | \, h(p) \, \rangle
  =2p^0 (2\pi)^3 \delta^{(3)} (\vec p-\vec p\, ')$,
its mass dimension is $[1/M]$.
If a hadron is made of $n_h$ elementary constituents,
its state vector could be written as 
\begin{align}
| \, h \, \rangle = \sqrt{N_h} \, | \, n_h \, \rangle , \ \ \ 
[\sqrt{N_h}] = [M^{n_h-1}]
\label{eqn:mass-counting}
\end{align}
where the second equation indicates that 
the normalization factor $\sqrt{N_h}$ has the mass
dimension $[M^{n_h-1}]$ if the state vector of each constituent 
has the mass dimension $[1/M]$.

%%%%%%%%%%%%%%%%%%%%%%%%
Here, we explain that the normalization factor $N_h$ 
is free from the hard momentum scale. 
Let us take a pion state as an example.
A pion state with momentum $p\simeq (p^+,0^-, \vec 0_T)$ 
in the c.m. system is expressed in terms of the Bethe-Salpeter (BS) 
wave function of the leading Fock state as   
\begin{align}
&
|\pi(p)\rangle=\int \frac{du}{\sqrt{u\bar{u}}}
\frac{d \vec k_T}{16\pi^3}
\Psi_{q\bar{q}/\pi}(u, \vec k_T)
|q(k_q)\bar{q}(k_{\bar{q}})\rangle+\cdots ,
\label{eqn:BS}
\end{align}
where $\bar{u}=1-u$ and the ellipses denote the higher Fock states, 
whose contribution to the exclusive scattering amplitude
is suppressed by some powers of $s$.
The leading Fock state consists of a quark and antiquark with momenta 
$k_q\simeq (up^+, \vec k_T^{\ 2}/(2up^+), \vec k_T)$ and 
$k_{\bar{q}}\simeq (\bar{u}p^+,\vec k_T^{\ 2}/(2\bar{u}p^+),-\vec k_T)$, 
respectively.   
Ignoring the higher Fock states, we have the normalization 
of the BS wave function given by 
\begin{align}
\int_0^1du\int \frac{d \vec k_T}{16\pi^3}
|\Psi_{q\bar{q}/\pi}(u,\vec k_T)|^2=1 .
\label{eqn:BS-2}
\end{align} 
Then, if one can assume that the wave function damps fast enough 
at large $|\vec k_T|$ such that it has non-zero values 
in the region $|\vec k_T|\lsim Q_{\rm had}$,  
where $Q_{\rm had}$ is the hadronic scale,  
its magnitude is given by 
$\Psi_{q\bar{q}/\pi}\simeq {\cal O}(1/Q_{had})$.
This means that the normalization factor is given by 
$\sqrt{N}_{\pi}\sim\int d \vec k_T
\Psi_{q\bar{q}/\pi}\simeq {\cal O}(Q_{had})$. 

Actually, in the perturbative calculation, the Feynman rule 
for the incoming pion, for example, is given from the following operator 
definition of the light-cone distribution amplitude as 
a matrix element of a bilocal operator 
between the pion and vacuum states \cite{erbl,chernyak-1984-summary}:
\begin{align}
&
\langle \, 0 \, | \, \overline d (0)_\alpha \, u (z)_\beta \,
                | \, \pi^+ (p) \, \rangle
\nonumber \\
& \ \ \ 
= \frac{i f_\pi}{4} \int_0^1 du \, 
% e^{-i (\bar u p^+ x^- +u p^+ y^-)}
e^{-i u p^+ z^-}
  \left ( \gamma_5 \pslash \right )_{\beta\alpha} \, \phi_\pi (u,\mu) ,
\label{eqn:qbar-q-matrix}
\end{align}
where $z=(0,z^-,\vec{0}_T)$ is a lightlike vector and 
$f_\pi$ is the pion decay constant defined as
$\langle \, 0 \, | \, \overline d (0) \gamma_\mu \gamma_5 \, u (0) \,
                | \, \pi (p) \, \rangle = i f_\pi p_\mu$, 
with the normalization $\int_0^1 du \, \phi_\pi (u)=1$.
A gauge link inserted between the two quark fields is understood on the LHS, 
so that the matrix element is gauge invariant.
The variables $u$ and $\mu$ are the longitudinal momentum fraction 
of a quark in the pion and the renormalization scale of the bilocal operator, 
respectively, where the latter dependence is governed by the ERBL evolution 
equation \cite{erbl}. 
The relation between the light-cone distribution amplitude and 
the BS wave function is given by \cite{mueller-1989}
\begin{align}
\int^{|\vec k_T|<\mu} \frac{d \vec k_T}{16\pi^3}
\Psi_{q\bar{q}/\pi}(u,\vec k_T)
\sim \frac{if_\pi}{4}\sqrt{\frac{2}{N_c}}\phi_\pi(u,\mu) ,
\label{eqn:BS-3}   
\end{align}
up to the scheme difference for subtracting the light-cone 
singularity in the bilocal operator. Here,
$N_c$ is the number of colors. From this expression, 
one can see the normalization factor is the order of magnitude of 
the pion decay constant: 
$\sqrt{N_\pi}\sim{\cal O}(f_\pi\simeq 0.13~{\rm GeV})$, 
which is the order of a typical hadron mass. 

The same discussions also apply for the nucleon.
By looking at its light-cone expression 
\cite{chernyak-1984-n}, one can explicitly see that
the normalization factor is of order of a soft mass scale squared.
Actually, the normalization factor is always given by the corresponding 
``decay constant" which is free from the hard momentum scale. 
%%%%%%%%%%%%%%%%%%%%%

Now, we consider the mass dimensions of 
the matrix element in Eq. (\ref{eqn:two-body-cross}) 
for obtaining the counting rule in the exclusive cross sections.
The scattering matrix $S$ is expressed by the transition matrix $T$ as
$S = \text{\boldmath{$1$}} + i(2\pi)^4 \delta^{(4)} (p_f-p_i) \, T$, 
so that the mass dimension of $T$ is $[T]=[M^4]$.
The matrix element $M_{ab \to cd}$ is given by $T$ as
\begin{align}
M_{ab \to cd} = \langle \, c d \, | \, T \, | \, a b \, \rangle 
              = \sqrt{N_a N_b N_c N_d}
                \langle \, n_c n_d \, | \, T \, | \, n_a n_b \, \rangle .
\end{align}
Because the normalization factors $N_{i}$ ($i=a,\, b,\, c,\, d$)
are expressed by soft constants, we consider the matrix element 
$\hat M_{ab \to cd}$ by excluding them, and then the remaining 
hard part should be expressed in terms of two variables $s$ and $t$:
\begin{align}
\hat M_{ab \to cd} \equiv \frac{1}{\sqrt{N_a N_b N_c N_d}} M_{ab \to cd}
& = \langle \, n_c n_d \, | \, T \, | \, n_a n_b \, \rangle
\nonumber \\
& \equiv \hat F_{ab \to cd} (s,t) .
\label{eqn:mhat}
\end{align}
From the dimensions $[T]=[M^4]$ and $[ \, |\, n_i \, \rangle \, ]=[1/M^{n_i}]$,
the dimension of the matrix element is given as
\begin{align}
[\hat M_{ab \to cd}] = [ \, 
       \langle \, n_c n_d \, | \, T \, | \, n_a n_b \, \rangle \, ]
= [M^{4-n}] , 
\end{align}
where $n \equiv n_a+n_b+n_c+n_d$.
The variable $s$ could be chosen as the only hard scale 
in the large-angle exclusive reaction, so that the matrix element is expressed,
by considering the mass dimension, as
\begin{align}
\hat M_{ab \to cd} = \hat F_{ab \to cd} (s,t)
= s^{(4-n)/2} F_{ab \to cd} (t/s) ,
\label{eqn:matrix-scaling}
\end{align}
where $F_{ab \to cd} (t/s)$ is a dimensionless quantity and
it is a function of scattering angle
$-2t/s = 1- \cos\theta_{cm}$ from Eq. (\ref{eqn:st}).
Using Eqs. (\ref{eqn:two-body-cross}), (\ref{eqn:mhat}),
and (\ref{eqn:matrix-scaling}), we obtain 
the constituent-counting expression for the cross section:
\begin{align}
\frac{d\sigma_{ab \to cd}}{dt} = \frac{1}{s^{n-2}} \, f_{ab \to cd}(t/s),
\label{eqn:cross-counting}
\end{align}
where $f(t/s)$ is the scattering-angle dependent part
multiplied by the normalization factors.
Because the mass dimensions of $f(t/s)$ are given by
$\left [ f(t/s) \right] 
= \left[ N_a N_b N_c N_d \, | F_{ab \to cd} |^2 \right]
= [ M^{2n-8} ]$,
the overall mass dimension of Eq. (\ref{eqn:cross-counting}) is, 
of course, $[1/M^4]$.
This is the derivation of the counting rule by considering the mass
dimensions.
Because it counts the number of constituents which actively participate
in the reaction, this scaling behavior is called 
the ``constituent-counting rule". 

%%%%%%%%%%%%%%%%%%%%%%%%%%%%%%%%%%%%%%%%%%%%%%%%%%%%%%%%%%%%%%%%%%%%%%%%%%%%%%%%
\subsection{Constituent-counting rule in perturbative QCD}\label{pqcd-counting}

The argument by the dimensional counting described above is intuitively clear, 
but it does not provide a ``proof" of the constituent counting rule. 
For example, Eq. (\ref{eqn:cross-counting}) is not valid for the contribution 
from disconnected so-called Landshoff diagrams \cite{landshoff-1974}.
%%%%%%%%%%%%%%%%%%%%%%%%%%%
Actually, each disconnected scattering amplitude is dimensionless,
while the condition that the separately scattered partons form the
hadrons in the final state requires that the c.m. momentum in each
subdiagram must coincide up to $Q_{\rm had}^{2}/s$.  For example, in
the elastic scattering of Fig. \ref{fig:elastic-ab-cd}, $x_a=x_b+{\cal
  O}(Q_{\rm had}^{2}/s)$ is imposed in the $[x]$-integral, which
eventually yields some powers of $Q_{\rm had}^{2}/s$
\cite{counting-Brodsky,exclusive-Brodsky}.  Such mechanism as the
origin of the scaling power is not included in the naive dimensional
counting in Sec. \ref{mass-counting}, where we treat $\sqrt{N_h}$ as a
dimensionful constant and assume that the $x$-integral does not affect
the dimensional counting.  Hence, in general, more rigorous arguments
based on perturbative calculations are needed to correctly identify
the scaling behavior \cite{counting-Brodsky, counting-2,
  Sivers-1982,exclusive-Brodsky, Mueller-Brodsky}.
%%%%%%%%%%%%%%%%%%%%%%%%%%%%
In this subsection, we discuss how the counting rule emerges in QCD 
from rough estimation of Feynman diagrams and possible complications.  

%%%%%%%%%%%%%%%%%%%%%%%%%%%% figure %%%%%%%%%%%%%%%%%%%%%%%%%%%%
\begin{figure}[b!]
\vspace{0.0cm}
\begin{center}
   \includegraphics[width=7.0cm]{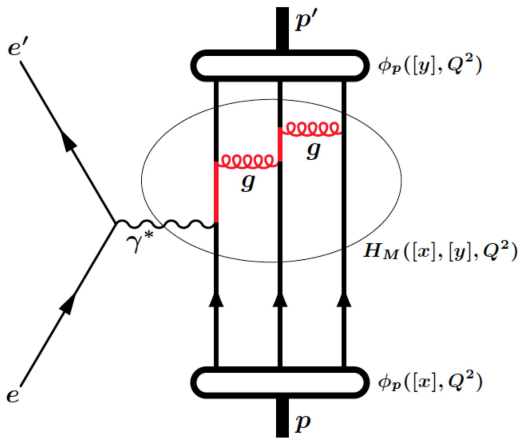}
\end{center}
\vspace{-0.5cm}
\caption{(Color online) A typical hard gluon-exchange process
in elastic electron-proton scattering ($e+p \to e'+p'$).
There are two hard quark propagators and two gluon ones
which contribute to the counting rule in the elastic form factor.}
\label{fig:ep-elastic-0}
\end{figure}
%%%%%%%%%%%%%%%%%%%%%%%%%%%% figure %%%%%%%%%%%%%%%%%%%%%%%%%%%%

Before stepping into an exclusive hadron-hadron reaction,
we explain a familiar elastic electron scattering
from the proton, $e+p \to e'+p'$. Its cross section is
described by elastic form factors of the proton:
\begin{align}
\! \! 
\langle \, p' \, | \, J^\mu \, | \, p \, \rangle
 = \overline u(p') \left [ 
 \gamma^\mu F_1 (Q^2) 
 + i \frac{\kappa}{2m_N} \sigma^{\mu\nu} q_\nu F_2 (Q^2) \right ] u (p)  ,
\label{eqn:form-factor-define}
\end{align}
where $F_1 (Q^2)$ and $F_2 (Q^2)$ are Dirac and Pauli form factors,
$\kappa$ is the anomalous magnetic moment, $m_N$ is the proton mass,
and $Q^2$ is given by the momentum of the virtual photon $q$
as $Q^2 = -q^2 \equiv \vec q^{\, 2}-(q^0)^2$. 
Then, the electric and magnetic form factors are defined by
$G_E = F_1 - \frac{\kappa Q^2}{4m_N^2} F_2$ and
$G_M = F_1 + \kappa F_2$.
In the following discussions, we consider 
the magnetic form factor $G_M$ which is dominant
in the cross section at large $Q^2$.

At large $Q^2$, the elastic form factor is factorized 
into a hard-scattering part $H_M$ and a soft part given by
the proton distribution amplitude $\phi_p$:
\begin{align}
\! \! \! \! 
G_M (Q^2) = \int \! [dx]  \int \! [dy] \,
    \phi_p ([y]) \,
% \nonumber \\ 
% & \ \ \ \ \ \ \ \ 
% \times
H_M ([x],[y],Q^2) \, \phi_p ([x]) ,
\label{eqn:form-factor-1}
\end{align}
where we suppress the scale dependence of $\phi_p$.
%%%%%
The hard amplitude $H_M ([x],[y],Q^2)$ should be evaluated in pQCD.
Because of the elastic scattering nature, the proton should not be
broken up by the large momentum given by the virtual photon
as shown in Fig. \ref{fig:ep-elastic-0}.
The only way to sustain the identity of the proton for
a given large momentum is to share the momentum among the constituents 
of the proton by exchanging hard gluons.
Therefore, the leading contribution to the elastic $ep$ cross
section should be described by the hard gluon exchange processes
typically shown in Fig. \ref{fig:ep-elastic-0}.

The amplitude $H_M$ is controlled by the momentum scale $Q$,
which is provided by the virtual photon,
in the two quark propagators and two gluon ones
in Fig. \ref{fig:ep-elastic-0}.
If we consider a frame with large momentum for the proton, 
specifically, the Breit frame where the virtual photon 4-momentum 
is given by $q=(0,\vec q \,)$, 
we have a relation
$ | \, \vec p \, |=| \, \vec {p}^{\, \prime} 
  \, | \equiv P \sim  {\cal O}(Q) \gg m_N$.
% , where $m_N$ is the mass of the proton.
There are additional hard factors due to each quark external line 
$u\sim \sqrt{P}$. 
More precisely, the three 
% incoming or outgoing 
quark lines are replaced by
$(\slash{p}\Gamma)_{\alpha\beta}(\Gamma^\prime u(p))_\gamma 
\sim (\sqrt{P})^{3}$,
where $\Gamma$ and $\Gamma'$ are appropriate $\gamma$ matrices, 
multiplied by the proton's distribution amplitudes 
\cite{chernyak-1984-n} for the incoming and outgoing proton. 
Anyway, there is a factor of 
$\sqrt{P} \sim \sqrt{Q}$ for each external quark line.
Therefore, there are two quark propagators $\sim 1/Q^2$,
two gluon propagators $\sim 1/(Q^2)^2$,
six external quark lines $\sim (\sqrt{Q} \,)^6$,
which give rise to the overall factor $1/(Q^{2})^{3/2}$:
\begin{align}
\langle \, p' \, | \, J^\mu \, | \, p \, \rangle
% H_M (y,z,Q^2) 
\sim \frac{1}{Q^2} \frac{\alpha_s (Q^2)^2}{(Q^2)^2} 
                 (\sqrt{Q} \,)^6 
            = \frac{\alpha_s (Q^2)^2}{(Q^2)^{3/2}} ,
\label{eqn:h-pqcd}
\end{align}
where $\alpha_s$ is the running coupling constant of QCD.

The proton distribution amplitude $\phi_p ([x])$ is the amplitude
for finding quarks with the momentum fractions $x_1$ and $x_2$
in the proton. This distribution amplitude also has  
a weak logarithmic $Q^2$ dependence \cite{erbl}
as we discussed in this section, which does not 
change the leading scaling behavior. 

There is one more factor which needs to be considered due to
the definition of the form factor in Eq. (\ref{eqn:form-factor-define}),
so that there is another hard factor 
$\overline u \gamma^\mu u \sim P \sim (Q^2)^{1/2}$
in front of the definition of the form factor.
Summarizing these results, we have
\begin{align}
\! \! 
G_M (Q^2) \sim \frac{1}{(Q^2)^{1/2}} 
\langle \, p' \, | \, J^\mu \, | \, p \, \rangle
\sim \frac{1}{Q^4} = \frac{1}{t^{n_N-1}} \ \ (n_N=3) ,
\label{eqn:f-pqcd}
\end{align}
where $t$ is the Mandelstam variable $t=-Q^2$ and $n_N=3$ is the number
of valence quarks in the proton. 
Actually, one can easily see that all factors of $Q$ cancel with each other 
except the ones from the $n_h-1$ gluon propagators.
Therefore,
the form factors generally scale as $1/t^{n_h-1}$, which is consistent with   
the constituent-counting rule in Eq. (\ref{eqn:cross-counting})
for the $e+h\rightarrow e+h$ scattering.
Such a scaling has been experimentally observed in the form factors
of the proton \cite{exp-form-factors}.

%%%%%%%%%%%%%%%%%%%%%%%%%%%% figure %%%%%%%%%%%%%%%%%%%%%%%%%%%%
\begin{figure}[b!]
\vspace{0.0cm}
\begin{center}
   \includegraphics[width=7.0cm]{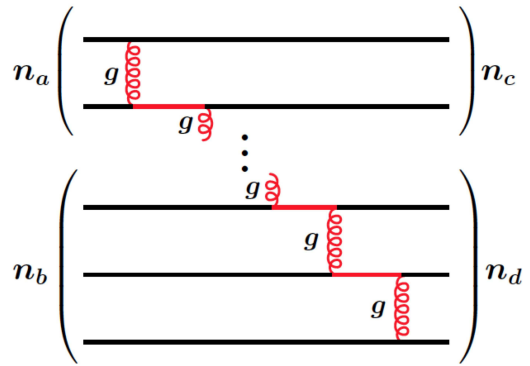}
\end{center}
\vspace{-0.5cm}
\caption{(Color online) Hard gluon exchange process for an exclusive
hadron-hadron reaction $a+b \to c+d$
with large momentum transfer.}
\label{fig:gluon-ex-exclusive}
\end{figure}
%%%%%%%%%%%%%%%%%%%%%%%%%%%% figure %%%%%%%%%%%%%%%%%%%%%%%%%%%%

From these discussions, we understand a scaling rule
for large-angle exclusive reactions in the following manner.
First, for finding the scaling behavior, it is enough to 
consider a Feynman diagram with the simplest topology as shown 
in Fig. \ref{fig:gluon-ex-exclusive}.
For the time being, we forget the flavor contents of the hadrons.
In order to become an exclusive reaction with
large momentum transfer, a hard gluon
should be exchanged between a quark in the hadron $a$ and 
a quark in $b$. Then, the large momentum should be shared
within the hadrons by exchanging hard gluons as shown 
in the figure. 
%%%%%
Denoting the hard momentum as $P$ in an exclusive reaction,
we have the following rule for calculating the scaling behavior
of the cross section:
\begin{itemize}
\vspace{-0.15cm}
\item[(1)] Feynman diagram \\
First, leading and connected Feynman diagrams are drawn for the exclusive 
process by connecting $n/2$ quark lines by gluons.
\vspace{-0.15cm}
\item[(2)] Gluon propagators \\
The factor $1/P^2$ is assigned for each gluon propagator.
Because there are $n/2-1$ gluon propagators in 
Fig. \ref{fig:gluon-ex-exclusive}, the overall factor 
is $1/(P^2)^{n/2-1}$.
\vspace{-0.15cm}
\item[(3)] Quark propagators \\
The factor $1/P$ is assigned for each quark propagator.
There are $n/2-2$ quark propagators, so that 
the overall factor becomes $1/P^{\, n/2-2}$.
\vspace{-0.15cm}
\item[(4)] External quarks \\
The factor $\sqrt{P}$ is assigned for each external quark.
Because there are $n$ quarks in the initial and final states
in total, the overall factor is $(\sqrt{P})^n$.
\end{itemize}
Then, the matrix element $\hat M_{ab \to cd}$ has the mass dimension
\begin{align}
\! \! \! \!
[\hat M_{ab \to cd}] 
= \left [ \frac{1}{(P^2)^{n/2-1}} \frac{1}{P^{n/2-2}} P^{n/2}
  \right ]
= \left [ \frac{1}{s^{n/2-2}} \right ] .
\label{eqn:m-pqcd}
\end{align}
Because the hadron distribution amplitudes $\phi_{a,\, b,\, c,\, d}$
have the weak logarithmic scale dependence, 
the leading contribution should come from the hard matrix element.
Then, the cross section is given by the constituent-counting expression
of Eq. (\ref{eqn:cross-counting}) by using
Eqs. (\ref{eqn:two-body-cross}), (\ref{eqn:mhat}),
and (\ref{eqn:m-pqcd}).
This is a diagrammatic explanation of the constituent-counting rule 
in perturbative QCD.

%%%%%%%%%%%%%%%%%%%%%%%%%%%%%%%%
There are theoretical complications which need to be considered 
for the counting rule 
\cite{counting-Brodsky,exclusive-Brodsky,Sivers-1982,Mueller-Brodsky}.
One is that the disconnected diagrams do not necessarily obey 
the counting rule as we explained before. 
Actually, they and some correction diagrams to them develop 
the ``pinch singularity", 
which occurs when a denominator of a gluon propagator vanishes  
inside the interval of the $x$-integration \cite{landshoff-1974}. 
After regularizing the linear divergence of the infrared origin, 
the $x$-integral around the pinch singularity gives a power of 
$\sqrt{s}/m$ in the matrix element, where $m$ is a quark mass.      
For example, such diagrams in meson-meson scattering scale as $s^{-5}$, 
instead of $s^{-6}$ by the counting rule \cite{counting-Brodsky}. 
However, the configuration at the pinch singularity is associated with
the elastic scattering of colored particles, and is subject to 
the Sudakov effects \cite{Sudakov}. 
Actually, it has been shown that the Sudakov effects shift the scaling power 
of the hadron-hadron scattering amplitude significantly, and the resulting 
``effective" scaling power is close to the one by the counting rule 
\cite{Mueller-Brodsky,Botts:1989kf}.

Furthermore, the endpoint singularity at $x \to 0 \text{ or } 1$ could
also affect the scaling behavior. At the endpoints, the momentum
transfer to one of the quarks becomes soft and the rules (2)-(4) for
calculating the scaling behavior do not apply.  A typical endpoint
singularity is given by an integral $\sim \int dx
\alpha_s(xQ^2)\frac{\phi_h(x,\cdots)}{x}$, so that the validity of the
pQCD description in Eq. (\ref{eqn:mab-cd}), let alone the counting
rule, depends on the nonperturbative endpoint behavior of the
distribution amplitude.  According to the conformal symmetry of QCD,
the distribution amplitudes are linear: $\phi_h(x)\sim x$ as $x\sim 0$
in the asymptotic limit \cite{Braun:1989iv}, as is known for the pion
distribution amplitude $\phi_\pi (x) = 6 x (1-x)$.  A numerical study
for the pion form factor with the conventional collinear factorization
like Eq. (\ref{eqn:mab-cd}) suggests that the pQCD description is
valid only at the very high energy \cite{Isgur:1984jm}. On the other
hand, a more elaborate study using the $k_t$ factorization formalism
tells that the effects of the Sudakov form factor provide a sufficient
suppression of the contribution from the endpoint region above
$Q\simeq 10\Lambda_{\rm QCD}$ \cite{Li-1992}.  Unfortunately, the
precise experimental tests of pQCD for exclusive hard processes are
still premature, but the recent BaBar and Belle data \cite{B-factory}
for the photon-pion transition form factor are not far from the pQCD
result \cite{counting-2,exclusive-Brodsky,erbl}.

Despite these theoretical complications,  
the constituent-counting rule seems to work well for hard exclusive reactions 
\cite{bnl-exclusive}, so that the above mentioned contributions 
from the pinch/endpoint singularities are not expected to change 
the rule to a significant amount. 
Actually, it seems that the counting rule applies even at the energy 
which is lower than the region where the leading power QCD description 
is considered to be valid.  

So far, we have ignored the hadron helicity in the exclusive
processes.  When the transverse momenta are integrated, only the
S-wave states are projected, unless $x\sim 0$ or $1$.  Since the QCD
interaction conserves the quark helicity up to the ${\cal O}(m^2/Q^2)$
effects, the total hadron helicity is also conserved to that accuracy:
$\lambda_a+\lambda_b=\lambda_c+\lambda_d$
\cite{exclusive-Brodsky,Ji:2003fw}.  In other words, the helicity
non-conserving processes are suppressed by a factor of $m^2/Q^2$ from
the scaling behavior given by Eq. (\ref{eqn:cross-counting}).  We also
note that the large-angle elastic scatterings, $\pi+p \ra K + \Lambda,
K + \Lambda(1405)$ which we discuss in this paper, are given by the
quark exchange diagrams.  Therefore, there appears no pinch
singularity for these processes.
%%%%%%%%%%%%%%%%%%%%%%%%%%%%%%%%

%%%%%%%%%%%%%%%%%%%%%%%%%%%%%%%%%%%%%%%%%%%%%%%%%%%%%%%%%%%%%%%%%%%%%%%%%%%%%%%%
\subsection{Internal structure of hadrons by counting rule}
\label{hadron-config}

%%%%%%%%%%%%%%%%%%%%%%%%%%%% figure %%%%%%%%%%%%%%%%%%%%%%%%%%%%
\begin{figure}[t!]
\vspace{0.0cm}
\begin{center}
   \includegraphics[width=8.0cm]{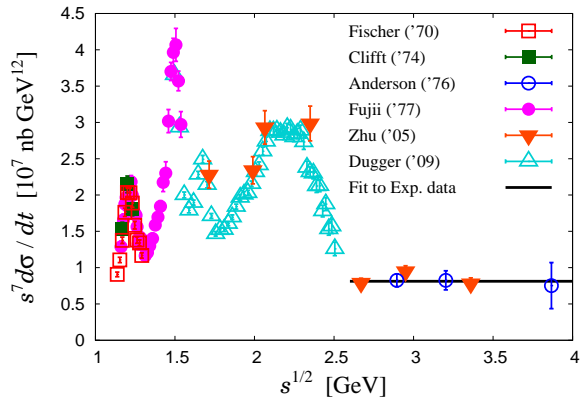}
\end{center}
\vspace{-0.8cm}
\caption{(Color online) $\gamma +p \to \pi^+ +n$ cross section
from resonances to a large momentum-transfer region.
The data are taken from Refs.
\cite{jlab-gamma-exclusive,gamma-exclusive-data}.
The straight line is a fit to the data at 
$\sqrt{s}>2.6$ GeV.}
\label{fig:jlab-gamma-p}
\end{figure}
%%%%%%%%%%%%%%%%%%%%%%%%%%%% figure %%%%%%%%%%%%%%%%%%%%%%%%%%%%

The scaling behavior of the exclusive cross section given by
the constituent-counting rule has been confirmed by a number of
experiments \cite{bnl-exclusive}. Another striking phenomenon, 
including the transition from hadron degrees of freedom 
to the quark degrees of freedom,
was observed by the reaction $\gamma +p \to \pi^+ +n$ 
in Fig. \ref{fig:jlab-gamma-p}. Here, the number of elementary 
constituents is $n=1+3+2+3=9$ in this reaction, and
the cross section is multiplied by the counting-rule factor $s^{9-2}$
in the ordinate, and it is shown as the function of the c.m.
energy $\sqrt{s}$. In the low-energy region $\sqrt{s} < 2.5$ GeV,
the cross section is described by contributions from nucleon
and delta resonances, whereas the scaling of $s^7 d\sigma /dt = const$
seems to be obtained at higher energy $\sqrt{s} > 2.6$ GeV.
Furthermore, the data suggest that the transition from
the hadron degrees of freedom to the quark ones occurs
at $\sqrt{s} \sim 2.5$ GeV, which is 1.6 GeV above
the proton mass.

We intend to use the counting rule for 
probing the internal structure of exotic hadron candidates. 
For example, ordinary $\Lambda$ should be counted as $n_\Lambda =3$; however,
it is expected to be $n_{\Lambda(1405)} =5$ if the structure is 
a five-quark configuration including
a $\bar KN$ molecule for $\Lambda \, (1405)$.
It is schematically shown in
Fig. \ref{fig:schematic-cross-section} 
by the cross section $s^8 d\sigma /dt$ at high energies.
If $\Lambda \, (1405)$ is a three-quark baryon,
it scales like $s^8 d\sigma /dt=$constant, whereas
it should be $s^8 d\sigma /dt \sim 1/s^2$
if $\Lambda \, (1405)$ is a five-quark state.

%%%%%%%%%%%%%%%%%%%%%%%%%%%% figure %%%%%%%%%%%%%%%%%%%%%%%%%%%%
\begin{figure}[h!]
\vspace{-0.5cm}
\begin{center}
   \includegraphics[width=10.0cm]{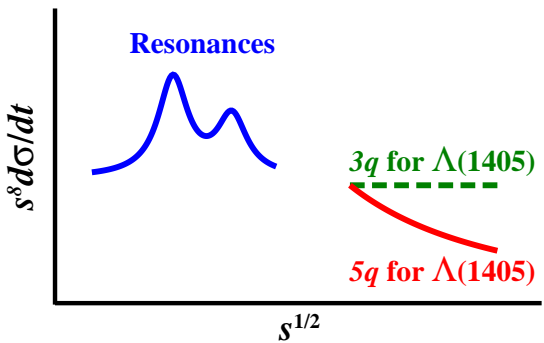}
\end{center}
\vspace{-0.9cm}
\caption{(Color online) 
Schematic figure of $\pi^- +p \to K^0 + \Lambda \, (1405)$ cross section
$s^8 d\sigma /dt$ as the function of $\sqrt{s}$
from the resonance region to the scaling one.
The scaling behavior at high energies indicates 
whether $\Lambda \, (1405)$ has an exotic five-quark configuration.}
\label{fig:schematic-cross-section}
\end{figure}
%%%%%%%%%%%%%%%%%%%%%%%%%%%% figure %%%%%%%%%%%%%%%%%%%%%%%%%%%%

\vspace{-0.0cm}
%%%%%%%%%%%%%%%%%%%%%%%%%%%%%%%%%%%%%%%%%%%%%%%%%%%%%%%%%%%%%%%%%%%%%%%%%%%%%%%%
%%%%%%%%%%%%%%%%%%%%%%%%%%%%%%%%%%%%%%%%%%%%%%%%%%%%%%%%%%%%%%%%%%%%%%%%%%%%%%%%
\section{Results}\label{results}

Our research purpose is to estimate the order of magnitude 
of the exclusive cross section of $\pi^- + p \to K^0 + \Lambda (1405)$
for future experimental proposals by considering existing 
experimental data and theoretical estimates to extend them 
to the large-momentum
transfer region, so that experimental measurements will be used 
for finding the internal structure of $\Lambda (1405)$ by the 
constituent-counting rule. As for the reference cross section, the order
of magnitude of the ground-state $\Lambda$ cross section is also
estimated from the data at high energies from the available data.
In addition, it is interesting to investigate the transition
from the hadron degrees of freedom to the quark ones,
as clearly shown in Fig. \ref{fig:jlab-gamma-p},
particularly for exotic hadrons.

At this stage, a successful theoretical description has not
been developed for estimating the magnitude of
exclusive cross sections in the perturbative QCD region 
\cite{pqcd-gamma-exclusive,duality-gamma-exclusive}
although the scaling behavior $d\sigma /dt \sim 1/s^{n-2}$ 
is well known. The following points need to be done
for the pQCD estimate.
First, there are many combinations of gluon-exchange processes
in addition to the typical example in Fig. \ref{fig:gluon-ex-exclusive}.
even for the ordinary three-quark $\Lambda$ and
especially if $\Lambda$ (1405) consists of five quarks.
The number of diagrams is significantly large, and 
they should be systematically calculated.
%%%%%
Second, the distribution amplitudes of hadrons
have not been determined, and they are necessary for calculating
the absolute cross section as obvious from Eq. (\ref{eqn:mab-cd}).
Even the distribution amplitude for the pion has not been 
established yet.
%%%%%
In spite of these issues, the counting rule is a valid
theoretical prediction in perturbative QCD and it could be
used for experimental studies on exotic hadrons.
For experimental proposals and actual measurements,
the order of magnitude of the $\Lambda (1405)$-production cross section
is needed. Therefore, we intend to provide such information in this work.

%%%%%%%%%%%%%%%%%%%%%%%%%%%%%%%%%%%%%%%%%%%%%%%%%%%%%%%%%%%%%%%%%%%%%%%%%%%%%%%%
\subsection{Cross section for $\pi^- + p \to K^0 + \Lambda$}
\label{Lambda}

%%%%%%%%%%%%%%%%%%%%%%%%%%%% figure %%%%%%%%%%%%%%%%%%%%%%%%%%%%
\begin{figure}[b!]
\setcounter{figure}{5}
\vspace{0.0cm}
\begin{center}
   \includegraphics[width=8.0cm]{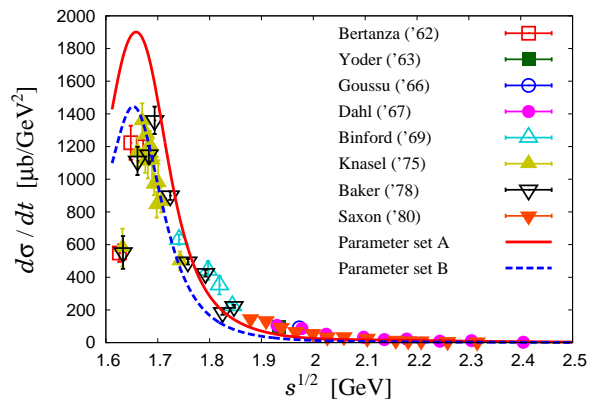}
\end{center}
\vspace{-0.5cm}
\caption{(Color online) 
Experimental data of $\pi^- +p \to K^0 +\Lambda$ cross section
$d\sigma /dt$ at $\theta_{cm}=90^\circ$
are compared with theoretical cross sections
calculated by the $N^*$ contributions
\cite{Ronchen:2012eg}.}
\label{fig:lambda-cross-exp-theo}
\end{figure}
%%%%%%%%%%%%%%%%%%%%%%%%%%%% figure %%%%%%%%%%%%%%%%%%%%%%%%%%%%

%%%%%%%%%%%%%%%%%%%%%%%%%%%% figure %%%%%%%%%%%%%%%%%%%%%%%%%%%%
\begin{figure*}[t!]
\setcounter{figure}{6}
\vspace{0.0cm}
\begin{center}
   \includegraphics[width=14.0cm]{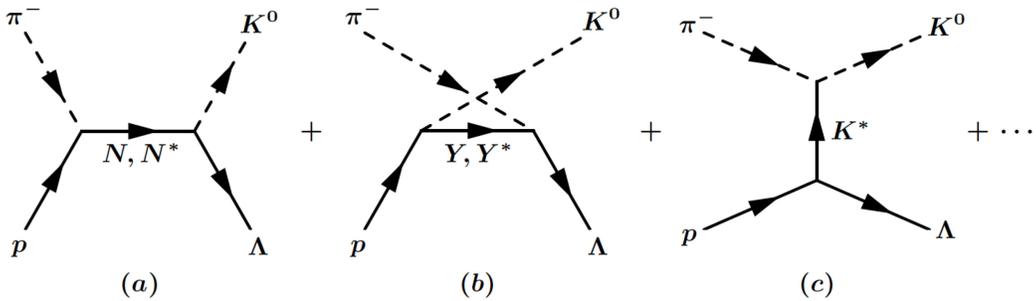}
\end{center}
\vspace{-0.5cm}
\caption{Subprocess for $\pi^- + p \to K^0 + \Lambda$
at low energies.}
\label{fig:lambda-prod}
\end{figure*}
%%%%%%%%%%%%%%%%%%%%%%%%%%%% figure %%%%%%%%%%%%%%%%%%%%%%%%%%%%

There are many available measurements on the cross section for $\pi^-
+ p \to K^0 + \Lambda$ \cite{lambda-prod-experiment,lambda-prod-data}
although the momentum transfer may not be sufficiently large.  We
could use these measurements together with the counting rule for
calculating the cross section in the large momentum-transfer region.
The cross-section measurements have been presented by $d\sigma
/d\Omega$ as the function of the c.m. scattering angle
$\theta_{cm}$. From them, we calculate ``experimental" cross sections
of $\pi^- + p \to K^0 + \Lambda$ at $\theta_{cm}=90^\circ$ as shown in
Fig. \ref{fig:jlab-gamma-p} as a function of the c.m. energy
$\sqrt{s}$. Since the measured values are not necessarily provided at
exactly $90^\circ$, we interpolate the data by smooth polynomials:
$d\sigma /d\Omega = \sum_{n=0}^{n_{max}} a_n (\cos\theta_{cm})^n$.
Then, the parameters $a_n$ are determined from the $\chi^2$ fit, and
the value at $\theta_{cm}=90^\circ$ is given by $d\sigma /d\Omega
|_{\theta_{cm}=90^\circ} = a_0$.  The results did not change
significantly as long as $n_{max}$ is taken as $n_{max} \sim 5$.  In
this work, only the statistical errors are included.  Then, the cross
sections are converted to $d\sigma /dt$ by changing the variable to
$t$.  The obtained data are shown in
Fig. \ref{fig:lambda-cross-exp-theo} together with theoretical
estimates with $N^{\ast}$ resonances \cite{Ronchen:2012eg} in order to
understand significant processes at low energies.

%%%%%%%%%%%%%%%%%%%%%%%%%%%% figure %%%%%%%%%%%%%%%%%%%%%%%%%%%%
\begin{figure}[b!]
\setcounter{figure}{7}
\vspace{0.0cm}
\begin{center}
   \includegraphics[width=8.0cm]{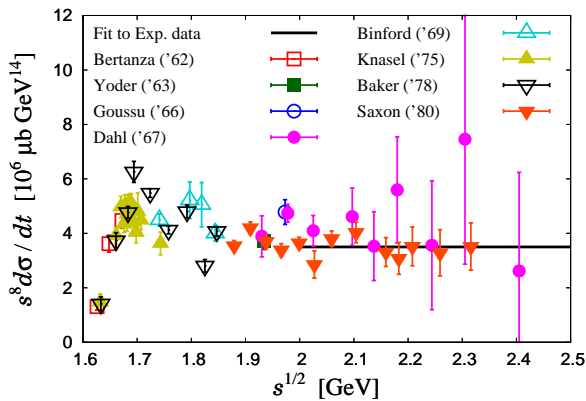}
\end{center}
\vspace{-0.5cm}
\caption{(Color online) 
Experimental data of $\pi^- +p \to K^0 +\Lambda$ cross section
$s^8 d\sigma /dt$ as the function of $\sqrt{s}$
\cite{lambda-prod-experiment}.
By considering the counting rule with $n=10$, the cross section
is multiplied by the factor $s^{n-2}$.
The line is a fit to the data at $\sqrt{s}>2$ GeV.}
\label{fig:lambda-cross-exp}
\end{figure}
%%%%%%%%%%%%%%%%%%%%%%%%%%%% figure %%%%%%%%%%%%%%%%%%%%%%%%%%%%

In Fig. \ref{fig:lambda-prod}, possible subprocesses are shown 
for the reaction $\pi^- + p \to K^0 + \Lambda$ at low energies 
by considering various intermediate resonances.
There were some studies on $\Lambda$ production processes
\cite{lambda-prod-theory-1}, and complete studies of 
hyperon-production reactions became available recently
by R\"onchen {\it et al.}~\cite{Ronchen:2012eg} 
and by Kamano {\it et al.}~\cite{lambda-prod-theory-3}.  
We do not step into the details of these reactions,
and simply the contributions from $s$-channel $N^*$
resonances in Fig. \ref{fig:lambda-prod}$(a)$ are
compared with the data in Fig. \ref{fig:lambda-cross-exp-theo}. 
As for the $N^*$, we took thirteen resonances:
$N(1535)$, $N(1650)$, $N(1440)$, $N(1710)$, $N(1750)$,
$N(1720)$, $N(1520)$, $N(1675)$, $N(1680)$, $N(1990)$,
$N(2190)$, $N(2250)$, and $N(2220)$.
Two possible parameter sets $A$ and $B$ are provided in 
Ref. \cite{Ronchen:2012eg} for these $N^*$ resonances,
and the two curves in Fig. \ref{fig:lambda-cross-exp-theo}
correspond to the two choices. 
At low energies, the experimental data
agree with the curves, which indicates that the dominant
subprocesses come from the intermediate $N^*$ resonances.
At higher energies at $\sqrt{s} > 1.8$ GeV, the curves
deviate from the data. It is because other processes, 
namely the crossed ones of $(b)$ and $t$-channel resonances
of $(c)$, and the coupled-channel effects 
contribute to the cross section.

%%%%%%%%%%%%%%%%%%%%%%%%%%%% figure %%%%%%%%%%%%%%%%%%%%%%%%%%%%
\begin{figure}[b!]
\setcounter{figure}{8}
\vspace{0.0cm}
\begin{center}
   \includegraphics[width=8.0cm]{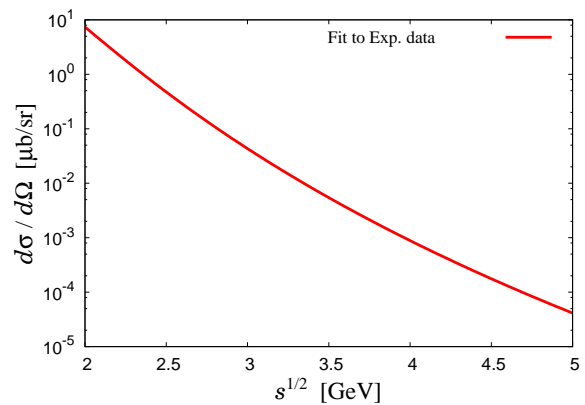}
\end{center}
\vspace{-0.5cm}
\caption{(Color online) 
The fitted cross section $d\sigma /dt$ 
of $\pi^- +p \to K^0 +\Lambda$ is extended to 
a high-energy region by assuming the constituent-counting rule
and the fitted value of Eq. (\ref{eqn:fitted-value}).}
\label{fig:lambda-cross-pqcd}
\end{figure}
%%%%%%%%%%%%%%%%%%%%%%%%%%%% figure %%%%%%%%%%%%%%%%%%%%%%%%%%%%

%%%%% This figure should be in the next subsection. %%%%%
%%%%%%%%%%%%%%%%%%%%%%%%%%%% figure %%%%%%%%%%%%%%%%%%%%%%%%%%%%
\begin{figure*}[t!]
\setcounter{figure}{9}
\begin{center}
  \includegraphics[scale=0.22]{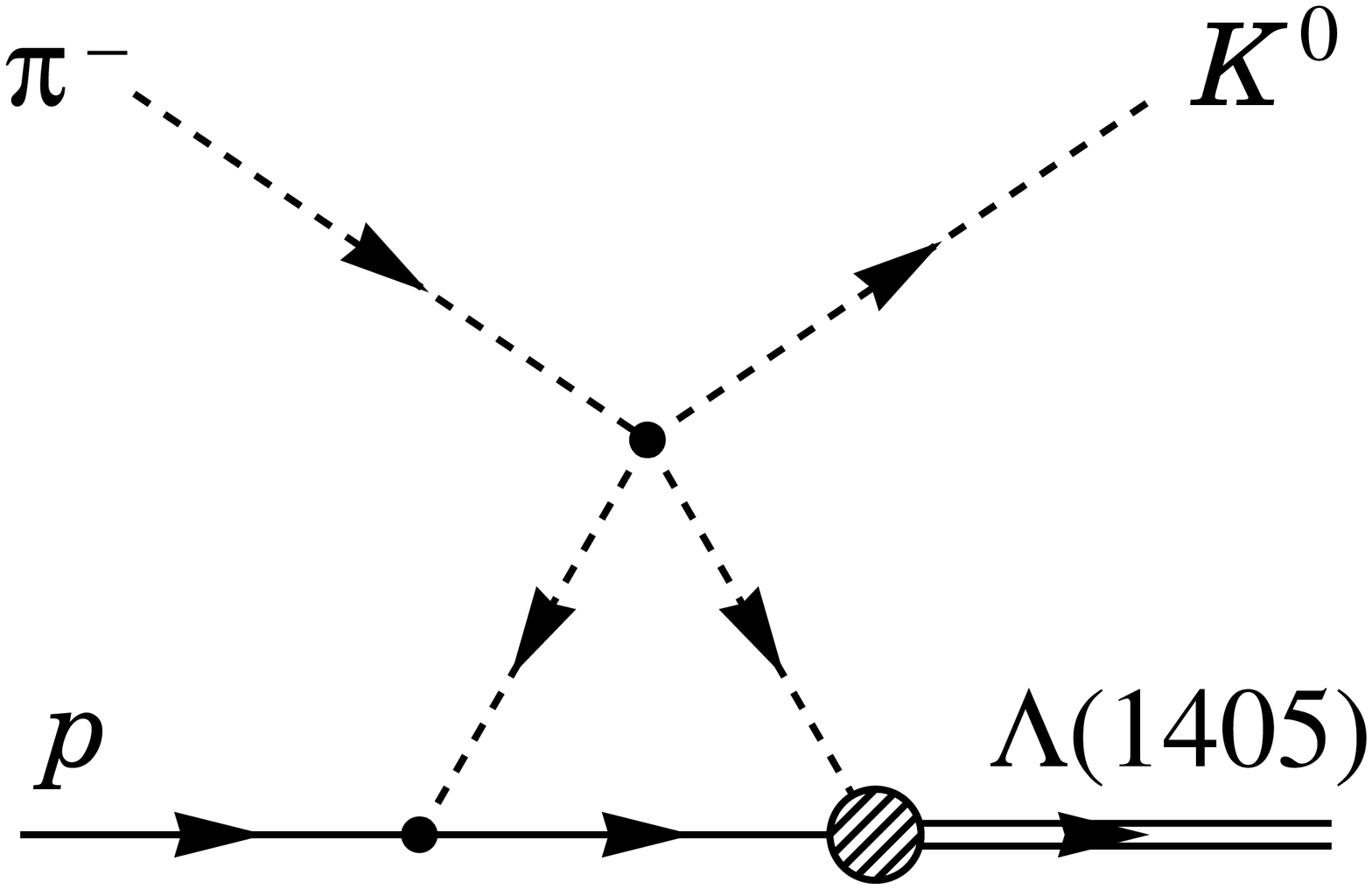}
  ~~~~
  \includegraphics[scale=0.22]{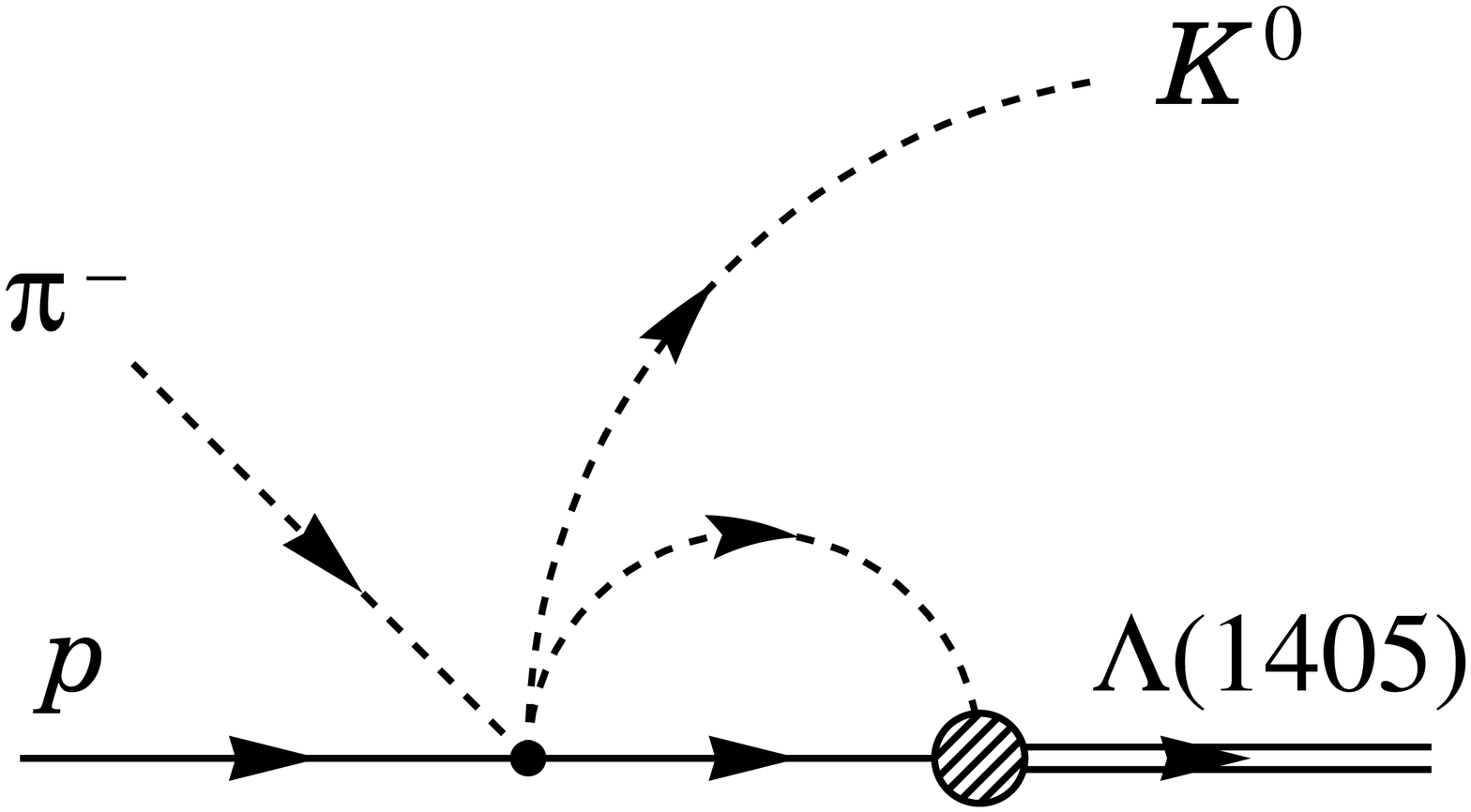}
  ~~~~
  \includegraphics[scale=0.22]{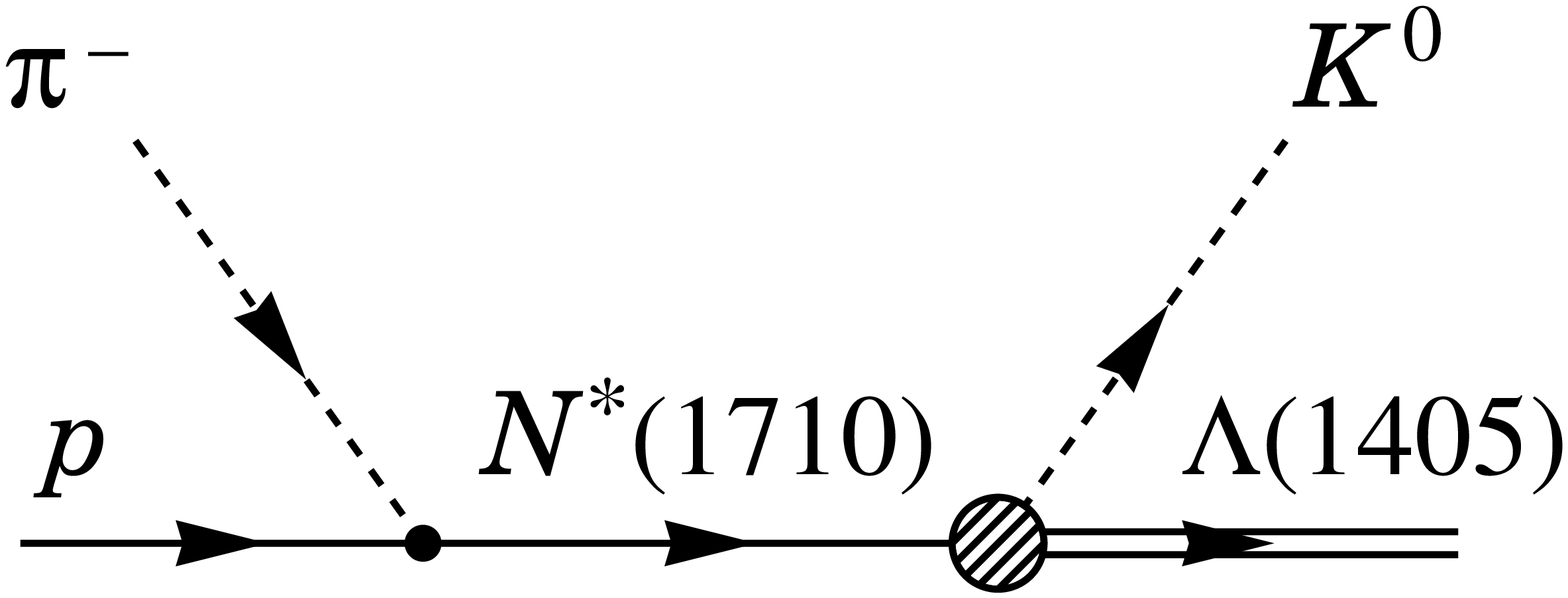}
  \caption{At low energies, the cross section of $\pi^- + p \to K^0 +
    \Lambda \, (1405)$ is calculated by the meson induced
    processes and the intermediate $s$-channel $N^*
    (1710)$~\cite{lambda-1405-prod-th}. }
\label{fig:lambda-1405-prod-1}
\end{center}
\end{figure*}
%%%%%%%%%%%%%%%%%%%%%%%%%%%% figure %%%%%%%%%%%%%%%%%%%%%%%%%%%%

It is, however, not obvious to predict cross section 
in the perturbative QCD region. 
The process should be described by Eq. (\ref{eqn:mab-cd})
at large-momentum transfer, but it is not possible
to obtain the accurate matrix element at this stage.
The hard part $H_{ab \to cd}$ could be calculated in perturbative QCD
in principle; however, there are too many processes 
to be evaluated easily by an analytical method. 
%%%%%
In addition,
the distribution amplitudes $\phi_{a,b,c,d}$ are not determined 
for $\pi$, $p$, $K$, and $\Lambda$ since there are still
discussions whether the functional form should be
the asymptotic form or the Chernyak-Zhitnitsky type 
even for the pion \cite{erbl, chernyak-1984-summary}
at present experimental energies.
In order to estimate the order of $\Lambda$ production
cross sections at high energies, we use a fit to the experimental data
in Fig. \ref{fig:lambda-cross-exp}.

In Fig. \ref{fig:lambda-cross-exp}, the experimental data
of $\pi^- + p \to K^0 + \Lambda$ are shown by the cross section
multiplied by $s^8$ which is the factor predicted by 
the constituent-counting rule with the total number $n=2+3+2+3=10$.
Bumpy resonance-like behavior is seen at low energies
$\sqrt{s} <1.9$ GeV, whereas the scaling appears at $\sqrt{s}>2$ GeV.
As explained in the last paragraph, it is not obvious what
should be the high-energy region where the perturbative QCD
can be applied. Therefore, we are not
confident whether the constant cross section at $\sqrt{s}>2$ GeV
indicates the scaling by the counting rule.
In the reaction $\gamma + p \to \pi^+ +n$ of Fig. \ref{fig:jlab-gamma-p},
the scaling starts from the excitation energy
$\sqrt{s}-m_p \simeq 2.5-0.9=1.6$ GeV. 
In Fig. \ref{fig:lambda-cross-exp}, it starts at
$\sqrt{s}-(m_K+m_\Lambda) \simeq 2.0-(0.5+1.1)=0.4$ GeV,
which is rather small in comparison with 
the $\gamma + p \to \pi^+ +n$ case. 
However, the hadron distribution amplitudes 
$\phi_{\pi,p,K^0,\Lambda}$ together with 
the hard scattering cross section $H_{\pi^- + p \to K^0 + \Lambda}$
are not known, so that there could be no wonder even if
the scaling starts from a lower energy.
In any case, we fit the experimental cross sections
at $\sqrt{s} >2$ GeV in Fig. \ref{fig:lambda-cross-exp}
by the straight line for an estimation in the scaling region.
From the fit to the experimental data, we obtain
\begin{align}
s^8 \frac{d\sigma}{dt} = (3.50 \pm 0.21)  
        \times 10^6 \ \mu b \ \text{GeV}^{14} .
\label{eqn:fitted-value}
\end{align}
On the other hand, fitting the experimental data 
at $\sqrt{s} > 2 \ \text{GeV}$ with the expression 
$d\sigma / dt = \text{(constant)}  \times s^{2-n}$, 
we obtain the scaling factor 
\begin{align}
n=10.1 \pm 0.6 ,
\end{align}
which is consistent with the three-quark structure for $\Lambda$.
It is an interesting and encouraging result for our studies.

Then, the cross section $d\sigma /d\Omega$ is shown in 
Fig. \ref{fig:lambda-cross-pqcd} for $\pi^- + p \to K^0 +\Lambda$
by extrapolating the constant cross-section value 
in Fig. \ref{fig:lambda-cross-exp} to the higher-energy region
up to $\sqrt{s} = 5$ GeV. The cross section is shown 
at $\theta_{cm}=90^\circ$ in the c.m. system.
Although it is a rough estimate, we show the cross section
for planning future experimental measurements
in comparison with the $\Lambda (1405)$ production
in Sec. \ref{Lambda-1405}.

%%%%%%%%%%%%%%%%%%%%%%%%%%%%%%%%%%%%%%%%%%%%%%%%%%%%%%%%%%%%%%%%%%%%%%%%%%%%%%%%
\subsection{Cross section for $\pi^- + p \to K^0 + \Lambda (1405)$}
\label{Lambda-1405}

We show the cross section of $\pi^- + p \to K^0 + \Lambda (1405)$ in
the same way as the $\Lambda$ production by using current information
from theoretical and experimental studies for finding the internal
structure of $\Lambda \, (1405)$ by the constituent-counting rule at high
energies.  However, both experimental and theoretical information is
very limited even in the resonance region for $\Lambda (1405)$.
Actually, there is only one
experiment for the pion induced $\Lambda (1405)$
production~\cite{lambda-1405-prod-ex},
and only the chiral unitary model~\cite{lambda-1405-prod-th}
is available for theoretical estimation.

%%%%%%%%%%%%%%%%%%%%%%%%%%%% figure %%%%%%%%%%%%%%%%%%%%%%%%%%%%
\begin{figure}[b!]
\hspace{0.0cm}
\begin{center}
  \includegraphics[width=8.0cm]{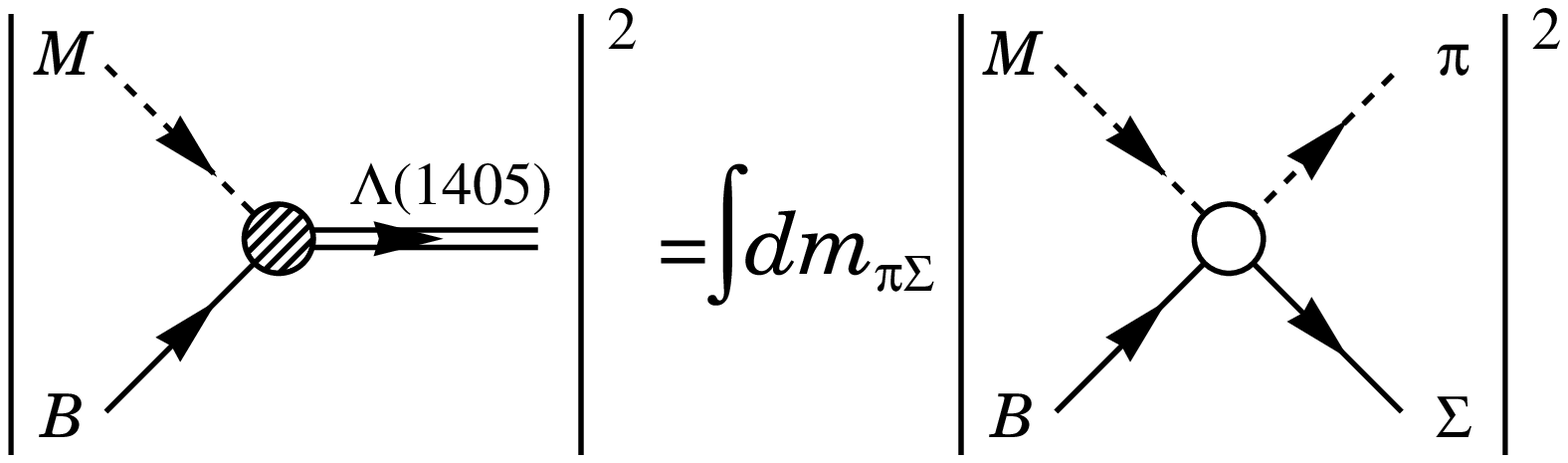}
\\
  \includegraphics[width=8.0cm]{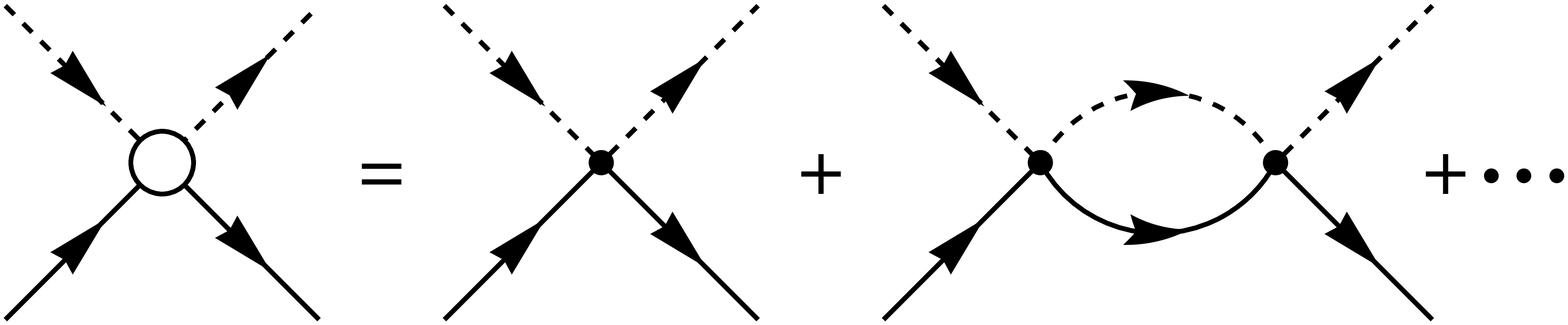}
\vspace{0.3cm}
\caption{The upper figures indicate that the cross section of 
  $\pi^- +p \to K^0 +\Lambda \, (1405)$ is calculated 
  by the cross section of
  $\pi^- +p \to K^0 +\pi +\Sigma$ integrated over the $\pi\Sigma$
  invariant mass $m_{\pi\Sigma}$ in the $\Lambda (1405)$ region.  The
  lower figures indicate that $\Lambda (1405)$ is generated in
  dynamical processes~\cite{lambda-1405-prod-th}.  The intermediate
  meson $M$ and baryon $B$ indicate ten sets of meson-baryon systems
  explained in the main text. }
\label{fig:lambda-1405-prod-2}
\end{center}
\end{figure}
%%%%%%%%%%%%%%%%%%%%%%%%%%%% figure %%%%%%%%%%%%%%%%%%%%%%%%%%%%
\vspace{-0.0cm}

In Ref.~\cite{lambda-1405-prod-th}, the pion induced $\Lambda (1405)$
production at low energies is theoretically studied by taking into
account the meson exchange contribution as well as the intermediate
$N^{\ast} (1710)$ $s$-channel formation as shown in 
Fig.~\ref{fig:lambda-1405-prod-1}. 
First, the cross section $\pi^- + p \to K^0 +\pi +\Sigma$ is calculated,
and then it is integrated over 
the invariant mass $m_{\pi\Sigma}$ of the final $\pi$ and $\Sigma$ 
in the $\Lambda (1405)$ energy region for obtaining the $\Lambda \,
(1405)$-production cross section as shown in Fig. \ref{fig:lambda-1405-prod-2}.
The couplings of $\pi + N \to N^{\ast} (1710)$ and $N^* (1710) \to K^0 +M +B$
are calculated from the $N^{\ast} (1710)$ partial decay widths 
with the flavor SU(3) symmetry.  Here,
the intermediate $MB$ states consist of the ten channels: $K^- p$,
$\bar K^0 n$, $\pi^0 \Lambda$, $\pi^0 \Sigma^0$, $\eta \Lambda$, $\eta
\Sigma^0$, $\pi^+ \Sigma^-$, $\pi^- \Sigma^+$, $K^+ \Xi^-$, and $K^0
\Xi^0$.

%%%%%%%%%%%%%%%%%%%%%%%%%%%% figure %%%%%%%%%%%%%%%%%%%%%%%%%%%%
\begin{figure}[b!]
\vspace{0.0cm}
\begin{center}
   \includegraphics[width=8.0cm]{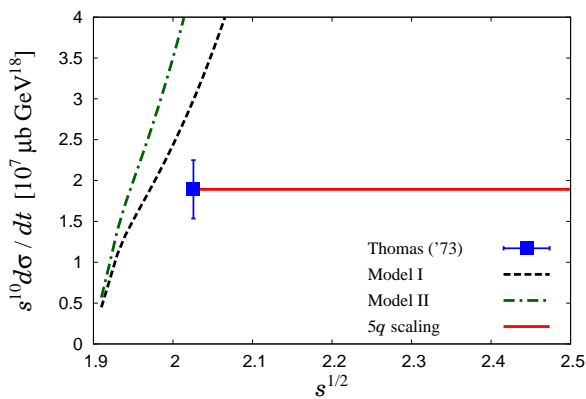}
\end{center}
\vspace{-0.5cm}
\caption{(Color online) 
Experimental data of $\pi^- +p \to K^0 +\Lambda \, (1405)$
cross section $s^{10} d\sigma /dt$ 
\cite{lambda-1405-prod-ex}.
By considering the counting rule with $n=12$, namely
a five-quark state for $\Lambda \, (1405)$, the cross section
is multiplied by the factor $s^{n-2}$.
The theoretical models I and II are from Ref. \cite{lambda-1405-prod-th}.}
\label{fig:lambda-cross-exp-1405}
\end{figure}
%%%%%%%%%%%%%%%%%%%%%%%%%%%% figure %%%%%%%%%%%%%%%%%%%%%%%%%%%%

If the $\Lambda \, (1405)$ is a five-quark state, the total number of
interacting elementary fields is $n=2+3+2+5=12$. The constituent-counting
rule indicates the scaling $s^{10} d\sigma /dt=$constant, so that the
cross section multiplied by $s^{10}$ is shown in
Fig. \ref{fig:lambda-cross-exp-1405} for $\pi^- + p \to K^0 +\Lambda \,
(1405)$ as the function of $\sqrt{s}$, in which the data together with
the available theoretical calculation are plotted.  The experimental
cross section at $\theta_{cm}=90^\circ$ is extracted from the
measurement \cite{lambda-1405-prod-ex} in the same with the $\Lambda$
cross section, and its value
\begin{align}
s^{10} \frac{d\sigma}{dt} = (1.89  \pm  0.36)  
        \times 10^7 \ \mu b \ \text{GeV}^{18} ,
\label{eqn:fitted-value-1405}
\end{align}
at $\sqrt{s} = 2.02 \text{ GeV}$ is plotted in
Fig. \ref{fig:lambda-cross-exp-1405}.  On the other hand, the
theoretical estimates roughly agree with the data,
but they diverge at large energies at $\sqrt{s} >2.1$ GeV simply
because the strong energy dependence of $s^{10}$ cannot be suppressed
by the contributions from Fig. \ref{fig:lambda-1405-prod-1}.
In any case, other resonances and $t$ channel contributions should be
taken into account for a precise description of the cross section, and
such hadronic models cannot be used at high energies.  In this
sense, we inevitably have to use experimental information for
estimating the cross section in the scaling region. The straight line
is drawn in Fig. \ref{fig:lambda-cross-exp-1405} by assuming the
scaling function for the five-quark type $\Lambda \, (1405)$.

%%%%%%%%%%%%%%%%%%%%%%%%%%%% figure %%%%%%%%%%%%%%%%%%%%%%%%%%%%
\begin{figure}[t!]
\vspace{0.0cm}
\begin{center}
   \includegraphics[width=8.0cm]{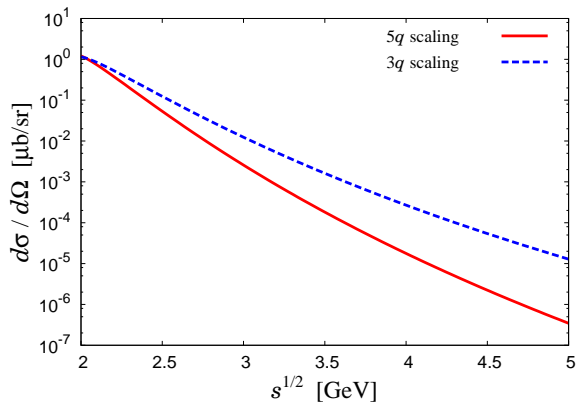}
\end{center}
\vspace{-0.5cm}
\caption{(Color online) The cross section $d\sigma /d \Omega$ of
  $\pi^- +p \to K^0 +\Lambda \, (1405)$ is extended to the high-energy
  region by assuming the constituent-counting rule and by using the
    experimental data in Eq. (\ref{eqn:fitted-value-1405}) at
    $\sqrt{s}=2.02 \text{ GeV}$~\cite{lambda-1405-prod-ex}.}
\label{fig:lambda-cross-pqcd-1405}
\end{figure}
%%%%%%%%%%%%%%%%%%%%%%%%%%%% figure %%%%%%%%%%%%%%%%%%%%%%%%%%%%

At this stage, the theoretical and experimental information is very
limited for estimating the order of the $\Lambda \, (1405)$ production
cross section at high energy.  On the other hand, even a rough
estimate of the cross section is needed for proposing a future
measurement at experimental facilities such as J-PARC. For this
purpose, we extended the cross section at $\sqrt{s}=2.02$ GeV to high
energies by assuming the scaling function with the five-quark $\Lambda
\, (1405)$.  Its cross section is shown in
Fig. \ref{fig:lambda-cross-pqcd-1405} by the solid curve with the
condition $d\sigma/d\Omega = 1.09 \pm 0.21 \ \mu b /sr$ at
$\sqrt{s}=2.02$ GeV.  In comparison, the dashed curve is also shown
for the scaling behavior to be observed if $\Lambda \, (1405)$ were an
ordinary three-quark baryon by assuming $s^{8} d\sigma /dt=$constant
and the same cross section at $\sqrt{s}=2.02$ GeV.  Because it is not
clear where the perturbative QCD region starts, the cross sections
should be considered as rough estimates.  In any case, there is a
distinct difference between the two functional forms if measurements
will be done at high energies.  In the scaling region, the quark-gluon
degrees of freedom explicitly appear, which results in the
constituent-counting rule, and the internal structure of $\Lambda \,
(1405)$ could be clarified.  If $\Lambda \, (1405)$ is a $\bar KN$
molecule, such investigations are similar to the scaling studies for
the deuteron \cite{j-parc-brosky} in the sense that both are bound
states of two hadrons.  Therefore, in this case $\Lambda \, (1405)$
can be treated simply as a five-quark state for studying the scaling
behavior.

%%%%%%%%%%%%%%%%%%%%%%%%%%%%%%%%%%%%%%%%%%%%%%%%%%%%%%%%%%%%%%%%%%%%%%%%%%%%%%%%
\subsection{Comments on experimental possibilities}
\label{future-experiments}

As for the future experimental measurements, there are possibilities
to measure $\pi^- + p \to K^0 + \Lambda (1405)$
at J-PARC by using the high-momentum beamline 
\cite{j-parc-workshop,J-PARC-exp-chang}
which will be ready in a few years.
There is also a high-momentum pion beam in the COMPASS experiment,
so that it could be possible.
%%%%%
Furthermore, there is a plan at LEPS 
(Laser Electron Photon beamline at SPring-8) II
to set up a detector for large-angle scattering measurements 
\cite{LEPS-future} in addition to the increase of photon energy.
Currently, the reaction $\gamma + p \to K^+ + \Lambda \, (1405)$
is taken up to the c.m. energy $\sqrt{s}=2.3$ GeV
within a limited scattering angle at LEPS,
and up to $\sqrt{s}=2.85$ GeV at JLab.
Then, the internal structure of $\Lambda \, (1405)$
could be also investigated by the exclusive reaction
$\gamma + p \to K^+ + \Lambda \, (1405)$
as we explained in this article.
According to the counting rule,
if the $\Lambda \, (1405)$ were an ordinary three-quark baryon,
the cross section should scale like 
$s^7 d\sigma/dt=$const as shown in Fig. \ref{fig:jlab-gamma-p}; 
however, it is $s^9 d\sigma/dt=$const if $\Lambda \, (1405)$
is a five-quark state.
Actually, there is an indication in Ref. \cite{lambda-1405-exp} 
that the $\Lambda \, (1405)$ photoproduction cross section is
suppressed at high energies in comparison with
the $\Sigma (1385)$ one.

Here, we discussed only $\Lambda \, (1405)$; however,
our idea can be used for investigating other exotic hadron
candidates by using the counting rule for exclusive reactions.
In addition to J-PARC and LEPS, there are several hadron and lepton 
beam facilities in the world, such as KEK-B, JLab, CERN-COMPASS, GSI, 
Fermilab, RHIC, LHC, etc. They could be used for such studies.
The idea of the counting rule is quite different from ordinary 
approaches at low energies, and we hope that our proposal will 
shed light on a new direction of exotic-hadron studies 
at high energies, where quark-gluon degrees of freedom 
appear. 

%%%%%%%%%%%%%%%%%%%%%%%%%%%%%%%%%%%%%%%%%%%%%%%%%%%%%%%%%%%%%%%%%%%%%%%%%%%%%%%%
%%%%%%%%%%%%%%%%%%%%%%%%%%%%%%%%%%%%%%%%%%%%%%%%%%%%%%%%%%%%%%%%%%%%%%%%%%%%%%%%
\section{Summary}\label{summary}

We proposed that the internal configuration of exotic hadron candidates
should be investigated by the scaling behavior given by the
constituent-counting rule for exclusive production processes.
As an example, the cross section was estimated for 
$\Lambda \, (1405)$ production processes $\pi^- +p \to K^0 +\Lambda \, (1405)$
together with the ground-state $\Lambda$ production
$\pi^- +p \to K^0 +\Lambda$.
The production cross sections were shown at $\theta_{cm}=90^\circ$
by using the existing experimental data and they were compared
with theoretical results in the resonance region. 
If the center-of-mass energy $\sqrt{s}$ becomes large enough,
the cross sections should be described by perturbative QCD
with light-cone wave functions of the hadrons.
The cross sections of this scaling region were simply estimated
by considering the counting rule in this work.
Depending on the quark configuration whether $\Lambda \, (1405)$
is a five-quark state (including $\bar K N$ molecule)
or ordinary three-quark hadron,
the scaling behavior is different. 
Measuring the exclusive cross sections at high energies,
we should be able to learn about the internal structure of
$\Lambda \, (1405)$.
This method is completely different from other studies 
at low energies and it provides a new approach for exotic-hadron 
studies by using high-energy processes.
We hope that our idea will be materialized as future measurements
at hadron facilities such as J-PARC and other facilities such as
LEPS and JLab.

\vspace{0.5cm} 

%%%%%%%%%%%%%%%%%%%%%%%%%%%%%%%%%%%%%%%%%%%%%%%%%%%%%%%%%%%%%%%%%%%%%%%%%%%%%%%%
%%%%%%%%%%%%%%%%%%%%%%%%%%%%%%%%%%%%%%%%%%%%%%%%%%%%%%%%%%%%%%%%%%%%%%%%%%%%%%%%
\begin{acknowledgements}
\vspace{-0.3cm}
The authors thank S. J. Brodsky,
W.-C. Chang, M. D\"oring, H. Gao, H. Kamano,
T. Mart, W. Melnitchouk, S. Sawada, and L. Zhu
for communications on exclusive processes, $\Lambda$ production 
processes, JLab data, and possible J-PARC measurements.
This work was partially supported by a Grant-in-Aid for Scientific 
Research on Priority Areas ``Elucidation of New Hadrons with a Variety 
of Flavors (E01: 21105006)" from the ministry of Education, Culture, 
Sports, Science and Technology of Japan.  
\end{acknowledgements}

\vspace{-0.0cm}
%%%%%%%%%%%%%%%%%%%%%%%%%%%%%%%%%%%%%%%%%%%%%%%%%%%%%%%%%%%%%%%%%%%%%%%%%%%%%%%%

%%%%%%%%%%%%%%%%%%%%%%%%%%%%%%%%%%%%%%%%%%%%%%%%%%%%%%%%%%%%%%%%%%%%%%%%%%%%%%%%

%%%%%%%%%%%%%%%%%%%%%%%%%%%%%%%%%%%%%%%%%%%%%%%%%%%%%%%%%%%%%%%%%%%%%%%%%%%%%%%%
\end{document}